\documentclass[%
 reprint,
superscriptaddress,
showpacs,
 amsmath,amssymb,
 aps,
 pra,
]{revtex4-1}

\usepackage{graphicx}
\usepackage{dcolumn}
\usepackage{bm}
\usepackage{epstopdf}
\usepackage{xcolor}

\usepackage{xspace}

\def \buildrelum#1\over#2{\mathrel{\mathop{\kern0pt #2}
                     \limits_{#1}   } }
\def \widetil{\mathrel\mathchar"0367}
\def \widesim{~~~~\lower.40em\hbox{$\widetil$}\mskip-10mu\succ ~~~}
\def \goes(#1){ \buildrelum #1 \over \longrightarrow}
\def \aslim(#1) {\buildrelum #1 \over \widesim}

\usepackage{dsfont}

\begin{document}


%
\title{Canonical quantization for quantum plasmonics with finite nanostructures}
\affiliation{Laboratoire Interdisciplinaire Carnot de Bourgogne, CNRS UMR 6303, Universit\'e Bourgogne Franche-Comt\'e,
BP 47870, 21078 Dijon, France}
\author{V. Dorier}
\affiliation{Laboratoire Interdisciplinaire Carnot de Bourgogne, CNRS UMR 6303, Universit\'e Bourgogne Franche-Comt\'e,
BP 47870, 21078 Dijon, France}
\author{J. Lampart}
\affiliation{Laboratoire Interdisciplinaire Carnot de Bourgogne, CNRS UMR 6303, Universit\'e Bourgogne Franche-Comt\'e,
BP 47870, 21078 Dijon, France}
\author{S. Gu\'erin}
\affiliation{Laboratoire Interdisciplinaire Carnot de Bourgogne, CNRS UMR 6303, Universit\'e Bourgogne Franche-Comt\'e,
BP 47870, 21078 Dijon, France}
\author{H. R. Jauslin}
\email{jauslin@u-bourgogne.fr}
\affiliation{Laboratoire Interdisciplinaire Carnot de Bourgogne, CNRS UMR 6303, Universit\'e Bourgogne Franche-Comt\'e,
BP 47870, 21078 Dijon, France}

\date{\today}

\begin{abstract}
The quantization of plasmons has been analyzed mostly under the assumption of an infinite-sized bulk medium interacting with the electromagnetic field. We reformulate it for finite-size media, such as metallic or dielectric nano-structures, highlighting sharp differences. By diagonalizing the Hamiltonian by means of a Lippmann-Schwinger equation, we
 show the contribution of two sets of bosonic operators, one stemming from medium fluctuations, and one from the electromagnetic field. 
The results apply to general models including dissipative and dispersive responses.
\end{abstract}
 


\maketitle
\section{Introduction}
Quantum effects in plasmonic systems are expected to open a whole branch of new perspectives both in fundamental physics and in technological applications
\cite{Tame-NatPhys-2013,Bozhevolnyi_case_for_quantum_plasmonics_2017,Dereux-Nature-2003, Altewischer-2002, DiMartino-2014, Dheur-2017, Novotny-Nano}, like e.g.~in quantum information, sensors, and generally in new types of circuits involving combinations of electronic and photonic effects. 
In the theories for quantum plasmonics there are two main approaches that have been widely used: (a) a phenomenological approach formulated in terms of quantum Langevin equations \cite{Welsch-phenomenological,Vogel-Welsch-book,Scheel-Buhmann-2009,Buhmann-book-Dispersion-forces-I} and (b) microscopic oscillator models for the medium coupled to the electromagnetic field \cite{Huttner-Barnett_1992,Suttorp-Wubs-2004-PRA,Suttorp-vanWonderen-2004-EPL,Philbin-2010-NJP}. The latter are variations of models of the type first proposed by Hopfield \cite{Hopfield_1958}. The quantization of a model for a homogeneous bulk medium was first treated by Huttner and Barnett \cite{Huttner-Barnett_1992}. Extensions to inhomogeneous media were treated in \cite{Suttorp-vanWonderen-2004-EPL, Philbin-2010-NJP}.
The main criterion for the choice of the microscopic model is that, if one integrates the equations for the medium and one inserts the obtained currents into the microscopic Maxwell equations, one should obtain the macroscopic Maxwell equations. The elementary quanta of the model are called \emph{plasmon-polaritons}.

Both approaches led to the following three basic relations:

[$A$] The diagonal plasmon-polariton Hamiltonian is of the form
\begin{align}
	\hat H=\int d^3r\int_0^\infty d\nu~\hbar\nu~\vec{\hat C}^\dag(\vec r,\nu)\cdot \vec{\hat C}(\vec r,\nu),\label{A}
\end{align}
where $\vec{\hat C}^\dag(\vec r,\nu)$ and $\vec{\hat C}(\vec r,\nu)$ are bosonic creation-annihilation operators.

[$B$] The electric field is expressed in terms of the bosonic operators as
\begin{align}
	\vec{\hat E}(\vec r)\!=\!\sqrt{\!\frac{\hbar\mu_0}{\pi c^2}}\!\!\!\int_0^\infty\!\!\! \!\!\!d\nu\!\!\int\! \!\!d^3r' ~\nu^2\epsilon_i^{1\!/\!2}(\vec r\,'\!,\!\nu)G(\vec r,\vec r\,'\!,\!\nu)\vec{\hat C}({\vec r}\,'\!,\!\nu)\!+\!\text{h.c.},\label{B}
\end{align}
where $\epsilon_i(\vec r\,',\nu)$ is the imaginary part of the dielectric coefficient, and $G(\vec r,\vec r\,',\nu)$ is a Green tensor satisfying
\begin{align}
	\left[\nabla\wedge\nabla\wedge - \epsilon(\vec{r},\nu)\frac{\nu^2}{c^2}\right]\bar{\bar{G}}(\vec r,\vec r\,',\nu)=\mathds{1}_3\delta(\vec r-\vec r\,').\label{green_eq}
\end{align}

[$C$] The Green tensor satisfies the following identity
\begin{align}
	\frac{\nu^2}{c^2}\!\!\int\!\!d^3r~\epsilon_i(\vec r)~G^T(\vec r,\vec r_A)G^*(\vec r,\vec r_B)=\mathfrak{Im}~G(\vec r_A,\vec r_B).\label{C}
\end{align}
These three relations may be justified in the case of a bulk medium, i.e.~when the dispersive and dissipative medium extends over all space $\mathbb{R}^3$ \cite{Welsch-phenomenological, Vogel-Welsch-book,Scheel-Buhmann-2009,Buhmann-book-Dispersion-forces-I, Suttorp-vanWonderen-2004-EPL, Philbin-2010-NJP}.

In this article we show that within the microscopic oscillator models neither of these three relations is valid for a finite medium, e.g.~in particular for a system with finite metallic or dielectric nano-structures. We present a procedure to construct the complete diagonalization of the Hamiltonian, Eq.~\eqref{ham3}, and the electric field operator, Eq.~\eqref{elec2},  adapted to finite media, which replace Eqs. \eqref{A} and \eqref{B}. 
Although we do not use the Green tensor \eqref{green_eq} explicitly in our construction, we remark that 
 for finite media a boundary term has to be added in Eq.~\eqref{C}, as described in \cite[Appendix A]{Drezet-PRA2017-equivalence}.

It has been stated in several instances in the literature \cite{DiStefano-2001,Drezet-PRA2017} that the formula \eqref{B} for the electric field observable cannot be generally correct, and that it has to be completed by including a term reflecting the contribution of the free electromagnetic field, while Eq. \eqref{B} describes only the electric field produced by the charges of the medium. A simple argument showing that \eqref{B} cannot be the complete formula for a finite medium is that if we take the limit where the size of the medium goes to zero, according to \eqref{B} the electric field observable would disappear. There must be another contribution that converges to the free electric field in the limit of vanishing medium.

Some extensions of Eq.~\eqref{B} were proposed in Refs.~\cite{DiStefano-2001,Drezet-PRA2017}, consisting in adding the free electric field
$\vec{E}_{\text{free}}$ :
\begin{align}
	\vec{\hat E}_{\text{tot}}=\vec{\hat E}_{\text{free}}+\vec{\hat E}_{\text{medium}},
\end{align}
with $\vec{\hat E}_{\text{medium}}$ given by \eqref{B}. This approach is not quite satisfactory, since $\vec{\hat E}_{\text{free}}$ was added without a justification from the diagonalization of the microscopic model. This might explain why this procedure was not widely adopted in the literature 
\cite{Khanbekyan-Knoll-Welsch}.

In this article, we show that the electric field operator has indeed two components: one that converges to the free electric field when the size of the medium goes to zero, and another one that vanishes in that limit. However, our result shows that the first component is not just the free electric field, but a modification induced by the presence of the medium. The second component is not equal to Eq.~\eqref{B}, but to an expression that is modified due to the presence of the first term.
We obtain this result by diagonalizing the microscopic model for a finite medium, as opposed to  an infinite bulk medium as in Refs.~\cite{Huttner-Barnett_1992,Suttorp-Wubs-2004-PRA,Suttorp-vanWonderen-2004-EPL,Philbin-2010-NJP}.
Furthermore, we show that the diagonalized Hamiltonian is not of the form \eqref{A}. It can be shown that, in the limit of vanishing coupling between the medium and the electromagnetic field, the operator \eqref{A} tends toward the Hamiltonian of the uncoupled medium, and the Hamiltonian of the uncoupled electromagnetic field would be missing. Finally, we compare in perturbation theory the electric field observable derived from our model with the one proposed in the literature.

\section{The microscopic plasmon-polariton model}
We consider the microscopic model for the classical electromagnetic field in interaction with a linear medium used in Ref.~\cite{Philbin-2010-NJP}. We consider a non-magnetic medium for simplicity, but our construction can easily be  extended to magnetic media and to the models of similar type considered e.g.~in Refs.~\cite{Suttorp-vanWonderen-2004-EPL,Suttorp-Wubs-2004-PRA}. It can be written in Hamiltonian form by choosing the following pairs of canonically conjugate variables: $\vec\Pi_{X}(\vec r,\nu)$, $\vec X(\vec r,\nu)$ for the harmonic oscillators describing the medium, the vector potential $\vec A(\vec r)$, and its canonically conjugate variable
\begin{align}
	\vec\Pi_A(\vec r):=-\vec D(\vec r)=-\epsilon_0\vec E(\vec r)-\!\int_0^\infty\!\!\! d\nu~\alpha(\vec r,\nu)\vec X(\vec r,\nu).\label{pia}
\end{align}
The Hamiltonian is
\begin{align}
	H=H_{\text{em}}+H_{\text{med}}+H_{\text{int}}\label{hsum}
\end{align}	
	with
	\begin{subequations}
	\label{hint}
\begin{align}	
	H_{\text{em}}&=\frac{1}{2}\int d^3 r\left[\frac{1}{\epsilon_0}\vec\Pi_A^2(\vec r)-\frac{1}{\mu_0}\vec A(\vec r)\cdot\Delta\vec A(\vec r)\right],\\
	H_{\text{med}}&=\frac{1}{2}\int_{V_m}\!\! d^3r\int_0^\infty \!\!d\nu \left[\vec\Pi^2_X(\vec r,\nu)+\nu^2\vec X^2(\vec r,\nu)\right],\\
	H_{\text{int}}&=\frac{1}{\epsilon_0}\int_{V_m} d^3 r~\vec\Pi_A(\vec r)\cdot\int_0^\infty d\nu~\alpha(\vec r,\nu)\vec X(\vec r,\nu)\nonumber\\
	&~~+\frac{1}{2\epsilon_0}\int_{V_m} d^3 r\left[\int_0^\infty d\nu~\alpha(\vec r,\nu)\vec X(\vec r,\nu)\right]^2,\label{Hint_3}
\end{align}
	\end{subequations}
where ${V_m}$ is the volume occupied by the medium, and $\alpha(\vec r,\nu)$ is the coupling. It is related to the imaginary part $\epsilon_i$
of the dielectric coefficient of the medium by $\alpha^2(\vec r,\nu)=2\epsilon_0\nu\epsilon_i(\vec r,\nu)/\pi$.
This microscopic model produces the macroscopic Maxwell equations when the medium variables are integrated \cite{Philbin-2010-NJP}.

We will use the following real basis of transverse generalized eigenfunctions of the Laplacian,
\begin{align}
	\vec\varphi_{\vec k,\sigma,\pi}(\vec r)=\begin{cases}
		\frac{1}{2\pi^{3/2}}\vec\varepsilon_\sigma(\vec k)\cos(\vec k\cdot\vec r),\quad \pi=c,\\[2mm]
		\frac{1}{2\pi^{3/2}}\vec\varepsilon_\sigma(\vec k)\sin(\vec k\cdot\vec r),\quad \pi=s,
		\end{cases}
\end{align}
where $\sigma=\pm$ is the index for the basis of linear polarization, and $\vec\varepsilon_\pm(\vec k)$ are two real unit vectors orthogonal to $\vec k$ and to each other.

In a first step, we apply the simple canonical transformation $(\vec\Pi_A,\vec A)\mapsto(\underline{p},\underline{q})$ 
\begin{subequations}
\begin{align} \label{Piphiq}
	\vec\Pi_A(\vec r)&=\int\!\! d\kappa\,\sqrt{\epsilon_0}\omega_\kappa\vec \varphi_\kappa(\vec r) q_\kappa,\\
	\vec A(\vec r)&=-\int\!\! d\kappa\, \frac{1}{\sqrt{\epsilon_0}\omega_\kappa}\vec\varphi_\kappa(\vec r)p_\kappa.
\end{align}
\end{subequations}
Here, we use the abbreviation for the labels $\kappa:=(\vec k,\sigma,\pi)$, and we denote by $\omega_\kappa=c|\vec k|$ the eigenfrequencies. The integral over $d\kappa$ is an abridged notation for a combination of integrals and sums.
	
In these variables, the coupled Hamiltonian takes the following form
\begin{align}
	H=\frac{1}{2}P\cdot P+\frac{1}{2}Q\cdot\Omega^2Q,\label{Hpq}
\end{align}
where
\begin{align}
	P:=\begin{bmatrix}
\underline{p}\\[1mm]
\vec\Pi_X
\end{bmatrix},\hspace{1cm}Q:=\begin{bmatrix}
\underline{q}\\[1mm]
\vec X
\end{bmatrix},\label{pq}
\end{align}
%
\begin{align}
	\Omega^2=\Omega_0^2+V.
\end{align}
$\Omega_0$ and $\Omega$ are called the frequency operators of the uncoupled and coupled systems, and $V$ is the coupling operator which can be explicitly extracted from Eq.~\eqref{Hint_3}. Since $V$ acts as a multiplication by $\alpha(\vec r,\nu)$ and an integration over the volume of the medium, it is uniquely defined for every dielectric coefficient(hence for every material) and for every geometry of the medium.

\section{General quantization procedure}
The quantization of a Hamiltonian of the general form \eqref{Hpq} and its expression in terms of creation-annihilation operators can be formulated in a systematic way \cite{deBievre_Where's_that_quantum,deBievre_Local_states_of_free_bose_fields} in terms of an orthonormal basis of generalized eigenfunctions of
\begin{align}
	\Omega^2\psi_{\lambda,d^\lambda}=\lambda^2\psi_{\lambda,d^\lambda},
\end{align}
where $d^\lambda$ is a label for the degeneracy indices. 
This quantization procedure can be thought off as a generalization of the usual quantization of the free electromagnetic field, in which the plane waves, which are eigenfunctions of the Laplacian,  are replaced by the eigenfunctions of $\Omega^2$.
The uncoupled  $\Omega_0^2$ has continuous spectrum. Under some  conditions on the coupling $V$ this is also the case for the coupled operator  $\Omega^2$. 
The diagonal quantized Hamiltonian can then be written as
\begin{align}
	\hat H=\int_0^\infty d\lambda\sum_{d^\lambda}\hbar\lambda\,\hat B^\dag_{\psi_{\lambda,d^\lambda}}\hat B_{\psi_{\lambda,d^\lambda}}.\label{ham1}
\end{align}
The notation $\sum_{d^\lambda}$ for the degeneracy is an abridged notation for a combination of sums and integrals.
%
The creation-annihilation operators act in the bosonic Fock space $\mathfrak{F}^B(\mathcal{P})$ \cite{Reed-Simon-vol1,deBievre_Where's_that_quantum,deBievre_Local_states_of_free_bose_fields}, with
\begin{align}
	\hat B^\dag_{\psi}|\varnothing\rangle=|\psi\rangle,
\end{align}
where $|\varnothing\rangle$ is the vacuum state and $\psi$ are functions corresponding to the coordinates of the classical phase space $\mathcal{P}$ that enters in the definition of the Fock space. Their interpretation is that $\hat B^\dag_{\psi}$ creates one quantum on the phase space mode $\psi$.
The orthonormality relation of the generalized eigenfunctions of $\Omega^2$ implies the canonical commutation relations.
The canonical observables can be expressed in terms of the creation-annihilation operators as
	\begin{subequations}
\begin{align}
	\hat Q&=\sqrt{\frac{\hbar}{2}}\int_0^\infty \!\!\!d\lambda\sum_{d^\lambda}\lambda^{-1/2}\left(\psi_{\lambda,d^\lambda}\hat B_{\psi_{\lambda,d^\lambda}}+\text{h.c.}\right),\\
	\hat P&=-i\sqrt{\frac{\hbar}{2}}\int_0^\infty \!\!\!d\lambda\sum_{d^\lambda}\lambda^{1/2}\left(\psi_{\lambda,d^\lambda}\hat B_{\psi_{\lambda,d^\lambda}}-\text{h.c.}\right).
\end{align}
	\end{subequations}

\section{Quantization of the plasmon-polariton model}
We can now apply this general procedure to the model \eqref{hsum}--\eqref{hint}. 
As we show below,  the (squared) frequency operator $\Omega^2$ has, in general, a continuous spectrum with two distinct families of generalized eigenfunctions, $\psi^e$ and $\psi^m$,
	\begin{subequations}
\begin{align}
	\Omega^2\psi^e_{\omega,d^\omega}&=\omega^2\psi^e_{\omega,d^\omega},\\
	\Omega^2\psi^m_{\nu,d^\nu}&=\nu^2\psi^m_{\nu,d^\nu},
\end{align}
	\end{subequations}
where the degeneracy indices are  $d^\omega:=(\vartheta,\phi,\sigma,\pi)$ with $\vartheta$ and $\eta$ the angles in spherical coordinates of $\vec k$, and $d^\nu:=(\vec r,j)$ with $j=1,2,3$ labeling the three components of the fields.

The eigenfunctions 
 have the same  structure as  Eq.~\eqref{pq}, i.e.~they are composed of two blocks:
\begin{align}
	\psi^e_{\omega,d^\omega}=\begin{bmatrix}
		u^e_{\omega,d^\omega}\\[1mm] v^{e}_{\omega,d^\omega}
	\end{bmatrix}, \qquad 
	\psi^m_{\nu,d^\nu}=\begin{bmatrix}
		u^m_{\nu,d^\nu}\\[1mm] v^{m}_{\nu,d^\nu}
	\end{bmatrix}.\label{structure_psi}
\end{align}
The functions $u$ in the upper block depend on the variables $(\vec k, \sigma, \pi)$ associated with the electromagnetic field, and the functions $\vec v$ in the lower block depend on the $(\nu, \vec r, j)$ of the oscillators.   
In the limit where the coupling $V$ vanishes, the $\psi^e$ tend to eigenfunctions of the free electromagnetic field and the $\psi^m$ tend to eigenfunctions of the uncoupled medium.
They satisfy the Lippmann-Schwinger equations~\cite[Sect XI.6, p.98]{Reed-Simon-vol3}:
	\begin{subequations}
	\label{Lippmann-Schwinger}
  \begin{eqnarray} \label{Lippmann-Schwinger-a}
 \psi^{e}_{\omega,d^\omega}  &=&\phi^{e}_{\omega,d^\omega} - \left(\Omega^2_0 -\omega^2 \mp i0^+ \right)^{-1} V  \psi^{e}_{\omega,d^\omega}, \\    \label{Lippmann-Schwinger-b}
  \psi^{m}_{\nu,d^\nu}  &=&\phi^{m}_{\nu,d^\nu} - \left(\Omega^2_0 -\nu^2 \mp i0^+ \right)^{-1} V  \psi^{m}_{\nu,d^\nu},
 \end{eqnarray}
	\end{subequations}
	with $\mp i0^+$ two possible ways to avoid the singularity in the complex plane. The functions $\phi$ are eigenfunctions of the uncoupled operator $\Omega_0^2$:
	\begin{subequations}
\begin{eqnarray}
	\phi^e_{\omega,d^\omega}(\omega',d^{\omega'})&=\begin{bmatrix}
	\delta(\omega-\omega')\delta_{d^\omega,d^{\omega'}}\\
	0
	\end{bmatrix},
	\end{eqnarray}
	corresponding to a plane wave in Fourier representation, and
	\begin{eqnarray}
	\phi^m_{\nu,\vec r,j}(\nu',\vec r\,',j')&=\begin{bmatrix}
		0\\
		\delta(\nu-\nu')\delta(\vec r-\vec r\,')\delta_{j,j'}
	\end{bmatrix},
\end{eqnarray}\label{phi}
	\end{subequations}
corresponding to an oscillator at position $\vec r$ oriented in direction $\vec e_j$.
This gives a one-to-one correspondence between the coupled and the uncoupled generalized eigenfunctions.
We will show below that the correspondence is unitary, which implies that the spectrum of $\Omega^2$ has the same degeneracy structure as the one of the uncoupled operator $\Omega_0^2$.

The diagonalized Hamiltonian takes the form
\begin{align} 
	\hat H=&\int_0^\infty\!\! d\omega\sum_{d^\omega}\hbar\omega~\hat D^\dag_{\omega,d^{\omega}}~\hat D_{\omega,d^{\omega}}\nonumber\\
	&+\!\int_0^\infty\!\!  d\nu \int_{V_m} \!\!d^3 r ~\hbar\nu~\vec{\hat C}^\dag_{\nu,\vec r}\cdot\vec{\hat C}_{\nu,\vec r},\label{ham3}
\end{align}
where we have introduced the notation
	\begin{subequations}
\begin{align}
\hat C_{\nu,\vec r,j}:=    &  \hat B_{\psi^m_{\nu,d^\nu}},  \\
\hat D_{\omega,d^{\omega}}:=    &  \hat B_{\psi^e_{\omega,d^\omega}}.
\end{align}
	\end{subequations}
This expression resembles the Hamiltonian \eqref{A} to which one would have added a free-field contribution. However, the creation-annihilation operators that appear in Eq.~\eqref{ham3} are a deformation of the ones of the uncoupled model.

%
In the exterior of the medium, the last term in Eq.~\eqref{pia} vanishes, and  Eq.~\eqref{Piphiq} gives
\begin{align}
	\vec{\hat E}(\vec r)=-\frac{1}{\sqrt{\epsilon_0}}
	\int\!\! d\kappa\, \omega_\kappa\vec\varphi_\kappa(\vec r)\hat q_\kappa.
\end{align}
Using Eq.~\eqref{pq}, the electric field observable 
can be written in terms of creation-annihilation operators on both sets of eigenmodes:
\begin{align}
	\vec{\hat E}(\vec r)&=\vec{\hat E}^e(\vec r)+\vec{\hat E}^m(\vec r),\label{elec2}\\
	\vec{\hat E}^e(\vec r)&=
	\int \!\! d\kappa \,
	\vec\beta_\kappa(\vec r)
	\int\limits_0^\infty\!\!\frac{d\omega}{\sqrt{\omega}}\sum_{d^\omega}\left(u^e_{\omega,d^\omega}\!(\kappa)\hat D_{\omega,d^\omega}+\text{h.c.}\right),\nonumber\\
	\vec{\hat E}^m(\vec r)&=
	\int \!\! d\kappa \,
	\vec\beta_\kappa(\vec r)
	\int\limits_0^\infty\!\!\frac{d\nu}{\sqrt{\nu}}\sum_{d^\nu}\left(u^m_{\nu,d^\nu}\!(\kappa)\hat C_{\nu,d^\nu}+\text{h.c.}\right),\nonumber
\end{align}
with $\vec\beta_\kappa(\vec r):=-\sqrt{\frac{\hbar}{2\epsilon_0}}\omega_\kappa\vec\varphi_\kappa(\vec r)$, and the sums over degeneracy indices correspond to the following combinations:
\begin{align}
	\sum_{d^\omega}&:=\frac{\omega^2}{c^3}\int_0^{\pi/2}d\vartheta\sin\vartheta\int_0^{2\pi}d\eta\sum_{\sigma=\pm}\sum_{\pi=c,s},\label{sum_kappa}\\
	\sum_{d^\nu}&:=\int_{V_m}d^3 r\sum_{j=1}^3.
\end{align}
Equation \eqref{elec2} is the main result of this construction. In particular, it exhibits the appropriate convergence in the limit of zero coupling.

In summary, the electric field observable at a point $\vec r$ in the exterior of the medium can be written according to Eq.~\eqref{elec2} as the sum of two terms. When the coupling goes to zero, one can show that the first term $\vec{\hat E}^e$ goes to the free electric field:
\begin{align}
	\vec{\hat{E}}^e(\vec r)\rightarrow -\!\int \!\!d\omega\sum_{d^\omega}\!\sqrt{\frac{\hbar\omega}{2\epsilon_0}}\vec\varphi_{\omega,d^\omega}(\vec r)\!\left[\hat{D}_{\omega,d^\omega}^0\!+\!\hat{D}_{\omega,d^\omega}^{0\dag}\right]\!\!,\label{Evac}
\end{align}
and the second term $\vec{\hat E}^m$ vanishes. However, for an arbitrary coupling, the first term is not equal to the free electric field, but to a deformation due to the presence of the medium. 

\section{Structure of the eigenfunctions}  
The uncoupled operator $\Omega_0^2$ has absolutely continuous spectrum of infinite degeneracy, labeled by the indices $d^\omega$ for the electromagnetic field and $d^\nu$ for the oscillators of the medium. We will now explain why the spectrum of the coupled operator $\Omega^2=\Omega_0^2+V$ has the same structure, including degeneracy, for the type of couplings that occur in plasmon-polariton models for a finite-size medium.

The methods and the intuition in the theory of continuous spectra are closely related to scattering theory. 
%
A central object in scattering theory is the M{\o}ller wave operator~\cite[Sect XI.3 p.18]{Reed-Simon-vol3}
\begin{equation}
 W_{\pm}(\Omega^2,\Omega_0^2)=\lim_{t\to \mp \infty} e^{i\Omega^2 t}e^{-i\Omega_0^2t}.
\end{equation}
If this operator is well defined (i.e.~if the strong limit $t\to \mp\infty$ exists) and is unitary  we have
\begin{equation}
 \Omega^2=W_{\pm} \Omega_0^2 W_{\pm}^\dagger,
\end{equation}
so $\Omega^2$ is unitarily equivalent to $\Omega_0^2$ and its spectrum must also be continuous with the same degeneracy. The generalized eigenfunctions are then given by $\psi_{\lambda,d^\lambda}=W_{\pm} \phi_{\lambda,d^\lambda}$. The unitarity of $W_{\pm}$ ensures that the eigenfunctions satisfy the usual relations of orthogonality and completeness. 

\begin{figure}[h!]
\centering
\includegraphics[width=.9\linewidth]{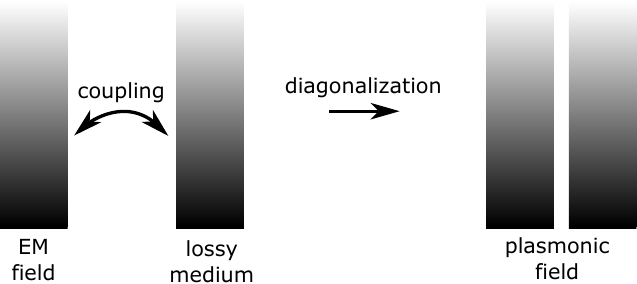}
\caption{Spectral structure of the plasmon-polariton model before and after diagonalization. The two continuous spectra structure is preserved by the unitarity of the diagonalization transformation.}\label{Fig}
\end{figure}

The coupling $V$ in the plasmon-polariton models belongs to a class that has been thoroughly studied since pioneering works by Friedrichs~\cite{Friedrichs-1937,Friedrichs-1948,Faddeev-2016}. 
The wave operator $W_{\pm}$ is an integral operator that can be constructed using the methods of~\cite{Yafaev-book-1992} (note~\cite{Note-Yafaev}).
The choice of sign of $W_\pm$ induces two possible sets of eigenfunctions verifying the Lippmann-Schwinger equation \eqref{Lippmann-Schwinger} either with $-i0^+$ (for $W_+$) or with $+i0^+$ (for $W_-$). The two sets of eigenfunctions are related by the unitary map $S=W_+^\dag W_-$, which is the scattering operator \cite{Reed-Simon-vol3}.
Further analysis reveals that $\Omega^2$ has no eigenvalues and purely absolutely continuous spectrum, provided the coupling function $\alpha(\nu)$ does not vanish for any $\nu>0$.
This assumption is satisfied for the usual models for dissipative and dispersive media.
Consequently, $W_{\pm}$ is unitary and this shows that the generalized eigenfunctions $\psi_{\lambda,d^\lambda}$ have the structure we claim. Figure~\ref{Fig} is a sketch of this preservation of the spectral structure.

\section{Comparison in perturbation theory}\label{Sec-comparison}
For a precise comparison between our construction and the results of Refs~\cite{Welsch-phenomenological,Vogel-Welsch-book,Scheel-Buhmann-2009,Buhmann-book-Dispersion-forces-I,Huttner-Barnett_1992,Suttorp-Wubs-2004-PRA,Suttorp-vanWonderen-2004-EPL,Philbin-2010-NJP,Hopfield_1958,DiStefano-2001,Drezet-PRA2017,Khanbekyan-Knoll-Welsch} one must be able to solve the Lippmann-Schwinger equations \eqref{Lippmann-Schwinger} in a given situation. A complete numerical solution which includes the high degree of degeneracy of the model will be the subject of future work. We present here a preliminary result where the equations \eqref{Lippmann-Schwinger} are solved in first order perturbation theory, i.e., where the volume of the medium and the dissipation are small. The details of the calculation are described in the appendix. Once the obtained functions $u^e$ and $u^m$ are inserted in the electric field operator \eqref{elec2}, it reads 
\begin{widetext}
\begin{align}
	\vec{\hat{E}}(\vec r)=\vec{\hat{E}}_0(\vec r)+\sqrt{\frac{\hbar\mu_0}{\pi c^2}}\int_0^\infty d\nu\int_{V_m}d^3 r'~\nu^2\epsilon_i^{1/2}(\nu,\vec r\,')\left[G_0^\pm(\nu;\vec r\,',\vec r)+F_{\text{nf}}(\nu;\vec r\,',\vec r)\right]\vec{\hat{C}}_{\nu,\vec r\,'}+h.c.,\label{E_G}
\end{align}
\end{widetext}
with $\vec r$ a position in the exterior of the medium. $\vec{\hat{E}}_0$ is of the same form as the electric field operator in vacuum~\eqref{Evac}. It is however not strictly the same, since the annihilation operator $\hat{D}$ acts on the plasmonic Fock space, which has a modified ground state. $G_0^\pm$ are two possible Green functions in vacuum:
\begin{align}
	G_0^\pm(\vec r\,',\vec r)=c^2\int d\vec k\sum_{\sigma,\pi}\frac{\vec\varphi_{\vec k,\sigma,\pi}(\vec r\,')\otimes\vec\varphi_{\vec k,\sigma,\pi}(\vec r)}{\omega^2-\nu^2\mp i0^+},
\end{align}
and $F_{\text{nf}}$ is a near-field term, which tends quickly to zero for $\vec r$ far from the surface of the medium:
\begin{align}
	F^{ij}_{\text{nf}}(\vec r\,'\!,\!\vec r)=\frac{-c^2}{4\pi\nu^2|\vec r-\vec r\,'|^3}\left(\delta_{ij}-\frac{3(r\,'_i-r_i)(r'_j-r_j)}{|\vec r-\vec r\,'|^2}\right).
\end{align}
The two possible Green tensors $G_0^{\pm}$ can be obtained for a given choice of the M\o ller operator $W_\pm$. One can show that $G_0^-$ satisfies the Sommerfeld radiation condition at infinity, with only outgoing waves to infinity, whereas $G_0^+$ contains only incoming waves from infinity. It is customary to use the Sommerfeld radiation condition. Both representations can be used, however, since they are linked by a unitary transformation.

The expression \eqref{E_G} is obtained at first perturbative order in $\sqrt{\epsilon_i}$. The link with Eq.~\eqref{B} can be done by also developping the Green equation \eqref{green_eq} to first order in $\sqrt{\epsilon_i}$, which corresponds to replacing the Green tensor in \eqref{B} by its vacuum expression $G_0$. We conclude that to first order in $\sqrt{\epsilon_i}$, the expression \eqref{E_G} takes the same form as the formulas proposed in Refs.~\cite{DiStefano-2001,Drezet-PRA2017} (i.e., an electric field of the form \eqref{B} but with an extra term associated with the free field). There is also an additional near-field term in \eqref{E_G}; this term diverges at the surface, and it is thus beyond the range of validity of perturbation theory.

It is worth mentioning that the ``free-field term'' $\vec{\hat{E}}_0$ is modified at higher orders of the perturbative development, which contradicts the idea that adding the contribution of the free field only is sufficient to achieve a complete quantization of the model.

\section{Conclusion and outlook.}  
In summary, unlike in Eq.~\eqref{A}, the diagonalized coupled Hamiltonian \eqref{ham3} is of the same form as the uncoupled one: it has absolutely continuous and positive spectrum with degeneracy labeled by the indices $d^\omega$ and $d^\nu$.

Since the starting point for the diagonalization procedure in Refs.~\cite{Suttorp-vanWonderen-2004-EPL, Philbin-2010-NJP} is the assumption of a diagonal Hamiltonian of the form \eqref{A}, the preceding arguments imply that the results of Refs.~\cite{Suttorp-vanWonderen-2004-EPL, Philbin-2010-NJP} cannot be applied to finite size media, in particular to nano-structures, where the Hamiltonian has the form \eqref{ham3}. The Fano-type diagonalization technique~\cite{Fano}, 
which was adapted in Refs.~\cite{Suttorp-vanWonderen-2004-EPL, Philbin-2010-NJP} was conceived initially to treat a discrete mode coupled to a continuum. A characteristic phenomenon is that, under some hypothesis, the discrete mode ``dissolves'' in the continuum, as was first proven by Friedrichs \cite{Friedrichs-1948} and rederived by Fano \cite{Fano}, and the coupled system has only continuous spectrum. This is an intuition that may have led to postulate a diagonal Hamiltonian of the form \eqref{A}. However, in the present models for finite media we have two coupled continua instead of a discrete mode and a continuum, and the Fano-Friedrichs mechanism does not apply. Instead we need an analysis of what happens when two continua interact, which is what we presented in this article.

We remark that the fact that the frequency operator $\Omega$ is diagonalized by a unitary transformation, given by the wave operator 
$ W_+$, guarantees that the operators 
$\vec{\hat C}_{\nu,\vec r}, \hat D_{\omega,d^\omega}$ 
automatically satisfy the bosonic commutation relations. In the method of diagonalization used in Refs.~\cite{Suttorp-vanWonderen-2004-EPL, Philbin-2010-NJP}, the commutation relations have to be imposed a posteriori, which is a technically non-trivial operation, in particular if one wants to extend it to include the two families of boson operators as in Eq.~\eqref{ham3}.

The relations \eqref{A}--\eqref{C} had been constructed implicitly for bulk systems, but  they have been widely used  to analyze phenomena in systems with finite size media, like e.g.~enhanced spontaneous emission, Purcell, Casimir and Polder effects, superradiant emission, among many others \cite{Buhmann-book-Dispersion-forces-I, applications}. These applications have to to be reassessed taking into account the modified perspective including the missing terms, e.g.~in the electric field. A natural question is whether the bulk formulas \eqref{A}--\eqref{C}
can be a good approximation to the complete expressions, in some particular regimes and beyond the perturbative approach described in Section~\ref{Sec-comparison}.

For applications it will be necessary to develop or adapt efficient numerical methods to solve the  Lippmann-Schwinger integral equations for media of different geometries (see e.g. Ref \cite{Atkinson-book-numerical-solution}), e.g.~to treat the fields of nano-antennas \cite{Lalanne-PRL-2013,Hughes-2019}. An analytical and numerical study of these equations and the consequences for spontaneous emission will appear elsewhere. 
The present results apply to a large class of open quantum systems coupled to a bath, featuring dispersion and dissipation.

\subsection*{Acknowledgments}
This work was supported by the ``Investissements d’Avenir'' program, project ISITE-BFC / IQUINS (ANR-15-IDEX-03), QUACO-PRC (ANR-17-CE40-0007-01) and the EUR-EIPHI Graduate School (17-EURE-0002). We. also acknowledge support from the European Union’s Horizon 2020 research and innovation program under the Marie Sklodowska-Curie grant agreement No. 765075 (LIMQUET)

\appendix

\section{First order perturbation theory}

We show here that a perturbation development to first order of the Lippmann-Schwinger equation results in the expression \eqref{E_G} of the electric field operator. To lighten the notation, we show it only for $G_0^-$ which satisfies the Sommerfeld radiation condition.

The Lippmann-Schwinger equations \eqref{Lippmann-Schwinger} take the general form
\begin{align}
	\psi_{\lambda,d^\lambda}^{e/m}=\phi_{\lambda,d^\lambda}^{e/m}-R_0 V\psi_{\lambda,d^\lambda}^{e/m},\label{LS}
\end{align}
where $R_0(\lambda)=(\Omega_0^2-\lambda^2+ i0^+)^{-1}.$ If the coupling $V$ is small, Eq.~\eqref{LS} can be developed as
\begin{align}
	\psi_{\lambda,d^\lambda}^{e/m}=\left[\mathds{1}-R_0 V+(R_0V)^2-\ldots\right]\phi_{\lambda,d^\lambda}^{e/m}.
\end{align}
In first order, we neglect all contributions of $V^n$ with $n>1$, we thus have
\begin{align}
	\psi_{\lambda,d^\lambda}^{e/m}=\phi_{\lambda,d^\lambda}^{e/m}-R_0 V\phi_{\lambda,d^\lambda}^{e/m}.\label{1storder}
\end{align}
How $V$ acts on a general vector of the Hilbert space is deduced from the Hamiltonian of interaction \eqref{Hint_3}. It takes the following form:
\begin{align}
	&V\begin{bmatrix}
		u(\kappa)\\[2mm] v(\nu,\vec r,j)
	\end{bmatrix}=\begin{bmatrix}
		(B v)(\kappa)\\[2mm] (B^Tu+Av)(\nu,\vec r,j)
	\end{bmatrix},\label{structure_V}
\end{align}
with
\begin{subequations}
\begin{align}
	&(B v)(\kappa)=\int\! d\nu \!\int_{V_m}\!\!d^3r~\omega\alpha(\nu,\vec r)\vec{\varphi}_\kappa(\vec r)\cdot\vec v(\nu,\vec r),\label{Bv}\\
	&(B^T u)(\nu,\vec r,j)=\alpha(\nu,\vec r)\int d\kappa~\omega\vec\varphi_\kappa(\vec r,j)u(\kappa),\\
	&(A v)(\nu,\vec r,j)=\alpha(\nu,\vec r)\int d\nu'\alpha(\nu',\vec r) v(\nu',\vec r,j),\label{Av}
\end{align}
\end{subequations}
where we recall the concise notation $\kappa=(\vec k,\sigma,\pi)$. Note that we absorbed $\epsilon_0$ by making the change of variable $\alpha\mapsto \sqrt{\epsilon_0}\alpha$. We then apply $V$ to the uncoupled eigenfunctions $\phi^{e/m}$ given by Eq.~\eqref{phi} and 
we insert it into Eq.~\eqref{1storder} to obtain the solutions of the Lippmann-Schwinger equations in a small coupling regime. They take the form \eqref{structure_psi} with:
\begin{subequations}
\begin{align}
	&u^e_\kappa(\kappa')=\delta(\kappa-\kappa'),\\[2mm]
	&v^e_\kappa(\nu,\vec r,j)=-\dfrac{\omega \varphi_{\kappa}(\vec r,j)\alpha(\nu,\vec r)}{\nu^2-\omega^2+ i0^+},\\[3mm]
	&u^m_{\nu,\vec r,j}(\kappa)=-\dfrac{\omega\varphi_{\kappa}(\vec r,j)\alpha(\nu,\vec r)}{\omega^2-\nu^2+ i0^+},\\[2mm]
	&v^m_{\nu,\vec r,j}(\nu'\!,\vec r\,'\!,j')=\delta(\nu-\nu')\delta(\vec r-\vec r\,')\delta_{j,j'}\nonumber\\
	&\hspace{2.5cm}-\dfrac{\alpha(\nu,\vec r)\alpha(\nu'\!,\vec r)}{\nu'^2-\nu^2+ i0^+}\delta(\vec r-\vec r\,')\delta_{j,j'}.
\end{align}
\end{subequations}
Now that the eigenfunctions of $\Omega^2$ have been found, we can use them to calculate the electric field in the exterior of the medium. We insert $u^e$ and $u^m$ into Eqs.~\eqref{elec2} and we obtain after some manipulations:
\begin{subequations}
\begin{align}
	\vec{\hat{E}}^e(\vec r)&=\vec{\hat{E}}^0(\vec r),\\
	\vec{\hat{E}}^m(\vec r)&=\sqrt{\frac{\hbar}{\pi\epsilon_0}}\!\int\! d\nu\!\int_{V_m}\!\!d^3r'\epsilon_i^{1/2}(\nu,\vec r\,')L(\nu;\vec r\,'\!,\vec r)\vec{\hat{C}}_{\nu,\vec r}\nonumber\\ &\hspace{.6\linewidth}+h.c.,\label{EmApp}
\end{align}
\end{subequations}
with
\begin{align}
	L(\vec r\,'\!,\vec r)=\!\int\!\! d^3 k\frac{\omega^2}{\omega^2-\nu^2+ i0^+}\sum_{\sigma,\pi}\vec\varphi_{\vec k,\sigma,\pi}(\vec r\,')\otimes\vec\varphi_{\vec k,\sigma,\pi}(\vec r).
\end{align}
We express the integrals in spherical coordinates with Eq.~\eqref{sum_kappa} and we use the identity
\begin{align}
	\frac{1}{\omega^2-\nu^2\mp i0^+}=\frac{\mathcal{P}}{\omega^2-\nu^2}\pm i\frac{\pi}{2\nu}\delta(\omega-\nu)
\end{align}
for $\omega,\nu>0$, which gives
\begin{align}
	L(\nu;\vec r\,'\!,\vec r)=\mathcal{P}\!\!\!\int_0^\infty \!\!\!d\omega\frac{\omega^2g(\omega;\vec r\,'\!,\vec r)}{\omega^2-\nu^2}- i\frac{\pi \nu}{2}g(\nu;\vec r\,'\!,\vec r)\label{EqL}
\end{align}
with
\begin{align}
	g(\omega;\vec r\,'\!,\vec r)=\sum_{d^\omega}\vec\varphi_{\omega,d^\omega}(\vec r\,')\otimes\vec\varphi_{\omega,d^\omega}(\vec r).\label{g}
\end{align}
The first term in \eqref{EqL} can be rewritten as
\begin{align*}
	&\mathcal{P}\!\!\!\int d\omega\left(\frac{\omega^2}{\omega^2-\nu^2}-1\right)g(\omega;\vec r\,',\vec r)+\int d\omega~ g(\omega;\vec r\,',\vec r)\\
	&=\mathcal{P}\!\!\!\int d\omega\frac{\nu^2g(\omega;\vec r\,',\vec r)}{\omega^2-\nu^2}+\int d\omega ~g(\omega;\vec r\,',\vec r).
\end{align*}
Because of Eq.~\eqref{g} and the completeness of the eigenfunctions $\vec \varphi_{\omega,d^\omega}$, we have
\begin{align*}
	\int d\omega ~g(\omega;\vec r\,',\vec r)=\delta_T(\vec r-\vec r\,'),
\end{align*}
with the transverse delta function defined as \cite{Stewart-2011}
\begin{align}
	\delta_{T}^{ij}&(\vec r-\vec r\,')=\frac{2}{3}\delta_{ij}\delta(\vec r-\vec r\,')\nonumber\\
	&-\frac{1}{4\pi|\vec r-\vec r\,'|^3}\left(\delta_{ij}-\frac{3(r_i-r_i')(r_j-r_j')}{|\vec r-\vec r\,'|^2}\right).
\end{align}
Since $\vec r\,'$ is defined only inside the medium whereas $\vec r$ is taken in the exterior, the first term in the RHS vanishes. Hence we obtain:
\begin{align}
	L(\nu;\vec r\,',\vec r)&=\nu^2\!\left[\int\! d\omega\frac{g(\omega;\vec r\,'\!,\vec r)}{\omega^2-\nu^2+ i0^+}+\frac{1}{c^2}F_{\text{nf}}(\nu;\vec r\,'\!,\vec r)\right]\!,\label{G}
\end{align}
where
\begin{align}
	F_{\text{nf}}^{ij}(\vec r\,'\!,\vec r):=\frac{-c^2}{4\pi\nu^2|\vec r-\vec r\,'|^3}\left(\delta_{ij}-\frac{3(r_i-r'_i)(r_j-r'_j)}{|\vec r-\vec r\,'|^2}\right).
\end{align}
In \eqref{G} we recognize the Green function in vacuum:
\begin{align}
	G_0^-(\nu;\vec r\,'\!,\vec r)&=c^2\!\int d\omega\frac{g(\omega;\vec r\,'\!,\vec r)}{\omega^2-\nu^2+ i0^+}\nonumber\\
	&\hspace{-.2cm}=c^2\!\int d^3k\sum_{\sigma,\pi}\frac{\vec \varphi_{\vec k,\sigma,\pi}(\vec r\,')\otimes\vec \varphi_{\vec k,\sigma,\pi}(\vec r)}{\omega^2-\nu^2+ i0^+},
\end{align}
hence
\begin{align}
	L=\frac{\nu^2}{c^2}[G_0^-+F_{\text{nf}}].
\end{align}
Introducing it back into Eq.~\eqref{EmApp} and using $\epsilon_0\mu_0c^2=1$, we obtain the electric field \eqref{E_G}.


\end{document}